\documentclass[12pt]{gen-p-l}

\newcommand{\p}{\partial}
\newcommand{\pb}{\bar \partial}

\newcommand{\cA}{\ensuremath{\mathcal{A}}}
\newcommand{\cD}{\ensuremath{\mathcal{D}}}
\newcommand{\cF}{\ensuremath{\mathcal{F}}}
\newcommand{\cH}{\ensuremath{\mathcal{H}}}
\newcommand{\cK}{\ensuremath{\mathcal{K}}}

\newcommand{\cN}{\ensuremath{\mathcal{N}}}
\newcommand{\cW}{\ensuremath{\mathcal{W}}}
\newcommand{\cZ}{\ensuremath{\mathcal{Z}}}

\newcommand{\Qb}{{\bar Q}}
\newcommand{\Xb}{{\bar X}}

\newcommand{\ib}{{\bar i}}
\newcommand{\jb}{{\bar j}}
\newcommand{\kb}{{\bar k}}
\newcommand{\lb}{{\bar l}}
\newcommand{\mb}{{\bar m}}

\newcommand{\Gammab}{{\bar \Gamma}}
\newcommand{\Lambdab}{\bar \Lambda}
\newcommand{\Phib}{{\bar \Phi}}

\newcommand{\alphab}{{\bar \alpha}}
\newcommand{\betab}{{\bar \beta}}
\newcommand{\gammab}{{\bar \gamma}}
\newcommand{\deltab}{{\bar \delta}}
\newcommand{\thetab}{{\bar \theta}}

\newcommand{\phib}{\ensuremath{\bar \phi}}
\newcommand{\psib}{\ensuremath{\bar \psi}}
\newcommand{\omegab}{{\bar \omega}}

\begin{document}


\title[On marginal deformations of general sigma-models]{On the marginal
  deformations of general (0,2) non-linear sigma-models}

\author{Ido Adam}
\address{Instituto de F\'isica Te\'orica,\\
Universidade Estadual Paulista\\
R.~Dr.~Bento~T.~Ferraz 271, Bloco II, 01140-070, S\~ao Paulo, SP,
Brasil\\
}
\email{idoadam.physics@yahoo.com}


\keywords{Sigma Models, Conformal Field Models in String Theory, Flux
  compactifications}

\begin{abstract}
  In this note we explore the possible marginal deformations
  of general (0,2) non-linear sigma-models, which arise as
  descriptions of the weakly-coupled (large radius) limits of
  four-dimensional $\cN = 1$ compactifications of the heterotic
  string, to lowest order in $\alpha'$ and first order in conformal
  perturbation theory. The results shed light from the world-sheet
  perspective on the classical moduli space of such
  compactifications. This is a contribution to the proceedings of
  String-Math 2012.
\end{abstract}

\thanks{IA's research was supported by FAPESP post-doctoral fellowship
  2010/07439-0.}

\maketitle

\section{Introduction}

One possible way of obtaining gauge theories in dimensions lower than
ten from superstring theory is to consider the heterotic string on
$\mathbb{R}^{d-1,1} \times M_{10-d}$, where $\mathbb{R}^{d-1,1}$ is
$d$-dimensional Minkowski space and $M_{10-d}$ is a compact
$(10-d)$-dimensional manifold. For such compactifications one has also
to specify the background gauge fields on $M_{10-d}$ (a vector bundle
$V$), which will break the large heterotic $\mathrm{SO(32)}$ or
$\mathrm{E}_8 \times \mathrm{E}_8$ gauge group into smaller gauge groups
more suitable for a realistic description of nature (for example
getting an $\mathrm{SU}(5)$ GUT).

There is particular interest in compactifications with $\cN = 1$
supersymmetry, since in that case there exist powerful
non-renormalization theorems protecting the superpotential from
perturbative $\alpha'$ corrections and in many cases ensuring the
existence of the vacuum in string perturbation theory. The
non-renormalization theorem can be violated by instanton effects, but
in some favorable cases these can be shown to be absent (see, for
example, \cite{Distler:1987ee,Silverstein:1995re,Beasley:2003fx}).

It is a well known result that $\cN = 1$ space-time supersymmetry in
four dimensions requires that the local (0,1) superconformal symmetry
of the world-sheet description of the Ramond-Neveu-Schwarz superstring
be enhanced to a global (0,2) superconformal symmetry (in which the
local one is embedded) and that extended space-time supersymmetry
leads to even higher superconformal symmetry on the world-sheet
\cite{Banks:1987cy,Banks:1988yz,Ferrara:1989bc}. When there is a
geometric description of the compactification (i.e., as a manifold and
vector bundle) the compact CFT on $M$ can be described in the large
radius limit as a (0,2) superconformal non-linear sigma-model. The
marginal deformations preserving the world-sheet supersymmetry (and
hence the target-space one) would then correspond to massless modes
parameterizing the moduli space of the compactification near the point
represented by the world-sheet non-linear sigma-model. This led the
authors of \cite{Melnikov:2011ez} to use an application of the methods
of \cite{Green:2010da} to unitary two-dimensional (0,2) SCFTs
\cite{Adam:2010unpublished} in order to determine those moduli in the
case of compactifications possessing $G = G' \times \mathrm{E}_8$
space-time gauge symmetry where $G'$ contains a non-anomalous
$\mathrm{U}(1)_\mathrm{L}$ symmetry.

In this publication we will examine to lowest order in $\alpha'$ and
to first order in conformal perturbation theory the general case of a
gauge group $G$ without requiring the existence of
$\mathrm{U}(1)_\mathrm{L}$ factors.

The paper is organized as follows. In section \ref{sec:general-NLSM}
we describe the general (0,2) non-linear sigma-model, in section
\ref{sec:marginal-deformations} we find the marginal deformations
preserving the (0,2) superconformal symmetry to lowest order in
$\alpha'$ and in conformal perturbation theory.

\section{The general (0,2) non-linear sigma-model} \label{sec:general-NLSM}

In this section the non-linear sigma-model describing the
weakly-coupled limit of a compactification of the heterotic string on
a complex manifold $M$ with a complex vector bundle $V \to M$ is
constructed. For the background to be consistent, it has to satisfy
the Bianchi identity $d\tilde H = \mathrm{ch}_2(\mathrm{T}M) -
\mathrm{ch}_2(V)$, where $\tilde H = dB - \omega(\mathrm{T}M) -
\omega(V)$ is the torsion field shifted by the Chern-Simons
three-forms constructed from the connection of the gauge vector bundle
and the tangent bundle. We will further assume that the
compactification has $\cN = 1$ target space supersymmetry, so the
sigma-model has (0,2) global supersymmetry
\cite{Banks:1987cy,Banks:1988yz,Ferrara:1989bc}. Target-space
supersymmetry to one-loop order also implies the Hermitian Yang-Mills
equations $\cF_{i j} = \cF_{\ib \jb} = 0$, $g^{\jb i} \cF_{i \jb} =
0$, where $\cF$ is the curvature of the bundle, so in particular, $V$
is a holomorphic vector bundle.

\subsection{(0,2) superspace conventions}

The easiest way to write the most-general (0,2) non-linear sigma-model
is in (0,2) superspace. We use mostly the conventions of
\cite{Witten:1993yc}. In particular the right-moving supercharges and
super-derivatives are
given by
\begin{equation}
  Q = \frac{\p}{\p \theta} + i \bar \theta \pb \ , \quad
  \bar Q = - \frac{\p}{\p \bar \theta} - i \theta \pb \ , \quad
  \cD = \frac{\p}{\p \theta} - i \bar \theta \pb \ , \quad
  \bar \cD = -\frac{\p}{\p \bar \theta} + i \theta \pb
\end{equation}

A generic (0,2) superfield is of the form
\begin{equation}
  \Phi = \phi + \sqrt{2} \theta \psi +\sqrt{2} \thetab \psib + i
  \theta \thetab F \ .
\end{equation}
A chiral superfield is constrained to satisfy $\bar \cD \Phi = 0$, while
an anti-chiral superfield satisfies $\cD \Phib = 0$. They are of the
form
\begin{equation}
  \Phi = \phi + \sqrt{2} \theta \psi - i \theta \thetab \pb \phi \ ,
  \quad
  \Phib = \phib - \sqrt{2} \thetab \psib + i \theta \thetab \pb \phib
  \ .
\end{equation}

The model also includes chiral and anti-chiral Fermi superfields
satisfying $\bar \cD \Gamma = 0$ and $\cD \Gammab = 0$. Their form is
\begin{equation}
  \Gamma = \gamma - \sqrt{2} \theta G - i \theta \thetab \pb \gamma
  \ , \quad
  \Gammab = \gammab - \sqrt{2} \thetab \bar G + i \theta \thetab \pb
  \gammab \ .
\end{equation}
The Hermiticity conditions relating the two kinds of fields are the
trivial ones $\theta^\dagger = \thetab$, $\phi^\dagger = \phib$,
$\psi^\dagger = \psib$ and $\gamma^\dagger = \gammab$.

\subsection{The (0,2) non-linear sigma-model}

The general (0,2) non-linear sigma-model has been written by
\cite{Dine:1986by}. Its field content is comprised of the chiral and
anti-chiral superfields $\Phi^i$, $\Phib^\ib$ ($i, \ib = 1,
\dots, n$), which are coordinates of a complex manifold $M$ of
dimension $n$ (for a Calabi-Yau three-fold, which is the case of most
interest, $n = 3$), and the chiral and anti-chiral Fermi superfields
$\Gamma^\alpha$ and $\Gammab^\alphab$, which take values in the
vector bundle $V$ over $M$. Their conformal weights and
$\mathrm{U}(1)_{\mathrm{R}}$ charges are given in
Table~\ref{tab:superfield-charges}.

The integrand of the superspace integral must be of conformal weight
(1,0) and have no $\mathrm{U}(1)_\mathrm{R}$ charge. The most general
such action is of the form
\begin{eqnarray}
  S & = & - \frac{i}{8 \pi \alpha'} \int d^2 x d^2 \theta \, \bigg[
    \cK_i(\Phi, \Phib) \p \Phi^i - \bar \cK_\ib (\Phi, \Phib) \p
    \Phib^\ib + \nonumber \\
    && {} + i \left( \cH_{\alpha \beta}(\Phi, \Phib) \Gamma^\alpha
    \Gamma^\beta + 2 \cH_{\alpha \betab}(\Phi, \Phib) \Gamma^\alpha
    \Gammab^\betab + \cH_{\alphab \betab}(\Phi, \Phib)
    \Gammab^\alphab \Gammab^\betab \right) \bigg] \ , \label{eq:NLSM-action}
\end{eqnarray} 
where $\cK_i(\Phi, \Phib)$ and $\bar \cK_\ib(\Phi, \Phib)$ can be regarded
as the (1,0)- and (0,1)-forms $\cK = \cK_i d\Phi^i$ and $\bar \cK =
\bar \cK_\ib d\Phib^\ib$ on the target manifold and $\cH_{\alpha
  \betab}$ is a Hermitian structure on the vector bundle.

\begin{table}
  \caption{Conformal weights and right R-charges of the various
    superfields}
  \label{tab:superfield-charges}
 \begin{tabular}{l|c|c|c}
    & $h$ & $\bar h$ & $\bar q$ \\
    \hline
    $\theta$ & 0 & $-\frac{1}{2}$ & $+1$ \\
    $\thetab$ & 0 & $-\frac{1}{2}$ & $-1$ \\
    $\cD$, $Q$ & $0$ & $\frac{1}{2}$ & $-1$ \\
    $\bar \cD$, $\Qb$ & $0$ & $\frac{1}{2}$ & $+1$ \\
    $\Phi^i$, $\Phib^\ib$ & $0$ & $0$ & $0$ \\
    $\Gamma^\alpha$, $\Gammab^\alphab$ & $\frac{1}{2}$ & $0$ & $0$
  \end{tabular}
\end{table}

In the absence of world-sheet boundaries, the action is invariant
under the transformations $\cK \to \cK + \p f$, $\bar \cK \to \bar \cK
+ \pb f$ for any (0,0)-form $f$, since these would shift the action by
a total divergence. Similarly, shifting $\cK \to \cK + \omega$, $\bar
\cK \to \bar \cK + \omegab$, where $\omega$ and $\omegab$ are a
holomorphic (1,0)-form and an anti-holomorphic (0,1)-form,
respectively, again only shifts the integral world-sheet integral by a
total derivative, because $\omega$ depends only on the
target-space-holomorphic fields $\Phi^i$ so
\begin{displaymath}
  \int_\Sigma \int d^2 \theta \, \omega(\Phi) = i \int_\Sigma
  \frac{\p}{\p \phi^i} \omega(\phi) \pb \phi^i = i \int_\Sigma \pb
  \omega(\phi) = 0 \ .
\end{displaymath}

Requiring that the action $S$ be real leads to the Hermiticity
conditions $\cK_i^\dagger = \bar \cK_\ib$, $\cH_{\alphab
  \betab}^\dagger = \cH_{\beta \alpha}$ and $\cH_{\alpha
  \betab}^\dagger = \cH_{\beta \alphab}$.

In order to get some intuition about the meaning of the various
background superfields, we also write the action in component form
(dropping total derivative terms which vanish in the absence of
world-sheet boundaries):
\begin{eqnarray}
  S & = & -\frac{1}{2 \pi \alpha'} \int d^2 z \, \Bigg[ \frac{1}{2} g_{i \jb}
    (\p \phi^i \pb \phib^\jb + \p \phib^\jb \pb \phi^i) + \frac{1}{2}
    B_{i \jb} (\p \phi^i \pb \phib^\jb - \p \phib^\jb \pb \phi^i)
    \nonumber \\
    && {} + i g_{i \jb} \psib^\jb \p \psi^i - i \psib^\jb \left(
    \Omega^-_{\jb k i} \p \phi^k + 
    \Omega^-_{\jb \kb i} \p \phib^\kb \right) \psi^i - i
    \gammab_\alpha (\pb \gamma^\alpha + \cA^\alpha_{i \gamma} \pb
    \phi^i \gamma^\gamma) \nonumber \\
    && {} -\frac{i}{2} \bar \cA^\alphab_{\ib
      \beta} \pb \phib^\ib \gamma_\alphab \gamma^\beta  - \frac{i}{2}
    \cA^\alpha_{i \betab} \pb \phi^i 
    \gammab_\alpha \gammab^\betab - \frac{1}{2} \tilde \cF^\alphab_{i
      \jb \beta} \psi^i \psib^\jb \gamma_\alphab \gamma^\beta + \tilde
    \cF^\alpha_{\jb i \beta} \psi^i \psib^\jb \gammab_\alpha
    \gamma^\beta \nonumber \\
    && {} + \frac{1}{2} \tilde \cF^\alpha_{\jb i \betab} \psi^i
    \psib^\jb \gammab_\alpha \gammab^\betab \Bigg] \ ,
\end{eqnarray}
where the metric and the $B$-fields are given by
\begin{equation}
   g_{i \jb} = \frac{1}{2} (\p_\jb \cK_i + \p_i \bar \cK_\jb) \ ,
   \quad
   B_{i \jb} = \frac{1}{2} (\p_\jb \cK_i - \p_i \bar \cK_\jb) \ ,
\end{equation}
and the torsion-twisted connection is
\begin{equation}
  \Omega^-_{\jb k i} = \Gamma_{\jb k i} - \frac{1}{2} H_{\jb k i} \ ,
  \quad
  \Omega^-_{\jb \kb i} = \Gamma_{\jb \kb i} - \frac{1}{2} H_{\jb \kb
    i} \ ,
\end{equation}
with $\Gamma$ being the Christoffel symbol of the first kind
associated with the metric $g_{i \jb}$ and $H = dB$ is the $H$-field.
Furthermore, for brevity we define the holomorphic and
anti-holomorphic connections on the vector bundle
\begin{equation}
  \cA^\alpha_{i \beta} \equiv \p_i \cH_{\beta \gammab} \cH^{\gammab
    \alpha} \ , \quad
  \bar \cA^\alphab_{\ib \betab} \equiv  \cH^{\alphab \gamma} \p_\ib
  \cH_{\gamma \betab}
\end{equation}
(we use the notation that $\cH^{\alphab \beta}$ is the inverse Hermitian metric
of $\cH_{\alpha \betab}$) as well as the symbols
\begin{eqnarray}
  \bar \cA^\alphab_{\ib \beta} & \equiv & \cH^{\alphab \gamma} \p_\ib
  \cH_{\gamma \beta} \ , \quad
  \cA^\alpha_{i \betab} \equiv \p_i \cH_{\betab \gammab} \cH^{\gammab
    \alpha} \ , \nonumber \\
  \tilde \cF^\alphab_{j \kb \beta} & \equiv & \p_j \bar
  \cA^\alphab_{\kb \beta} - \cA^\gamma_{j \beta} \bar
  \cA^\alphab_{\kb \gamma} \ , \quad
  \tilde \cF^\alpha_{\kb j \beta} \equiv \p_\kb \cA^\alpha_{j \beta} -
  \cA^\alpha_{j \gammab} \bar \cA^\gammab_{\kb \beta} \ ,
  \nonumber \\
  \tilde \cF^\alpha_{\kb j \betab} & \equiv & \p_\kb \cA^\alpha_{j \betab}
  - \bar \cA^\gammab_{\kb \betab} \cA^\alpha_{j \gammab} \ .
\end{eqnarray}
Note that these are not connections and curvatures of the vector
bundle (the curvature is $\cF^\alpha_{\jb i \beta} = \p_\jb
\cA^\alpha_{i \beta}$). We hope that the use of the letters $\cA$ and
$\cF$ will not cause any confusion.

In the sequel we will require the equations of motion for the various
superfields. Since the fields are either chiral or anti-chiral, we
need their variations to obey the chirality/anti-chirality
constraints. This is easily done by writing the variations as
\begin{equation}
  \delta \Phi^i = \bar \cD \delta X^i \ , \quad
  \delta \Phib^\ib = \cD \delta \Xb^\ib \ , \quad
  \delta \Gamma^\alpha = \bar \cD \delta \Lambda^\alpha \ , \quad
  \delta \Gammab^\alphab = \cD \delta \Lambdab^\alphab \ .
\end{equation}
The equations of motion thus obtained are
\begin{eqnarray}
  E^\Phi_i & = & 2 H_{j \kb i} \bar \cD \Phib^\kb \p \Phi^j - 2 \bar
  \cD (g_{i \jb} \p \Phib^\jb) + i \bar \cD \p_i \cH_{\alpha \beta}
  \Gamma^\alpha \Gamma^\beta - \nonumber \\
  && {} - 2 i \bar \cD (\p_i \cH_{\alpha \betab}
  \Gammab^\betab) \Gamma^\alpha+ i \bar \cD (\p_i \cH_{\alphab \betab}
  \Gammab^\alphab  \Gammab^\betab) = 0 \ , \label{eq:undeformed-eom-1} \\
  E^\Phib_\ib & = & 2 \cD (g_{j \ib} \p \Phi^j) + 2 H_{k \jb \ib} \cD
  \Phi^b \p \Phib^\jb + i \cD (\p_\ib \cH_{\alpha \beta} \Gamma^\alpha
  \Gamma^\beta) + \nonumber \\
  && {} + 2 i \cD (\p_\ib \cH_{\alpha \betab} \Gamma^\alpha
  \Gammab^\betab) + 
  i \cD \p_\ib \cH_{\alphab \betab} \Gammab^\alphab
  \Gammab^\betab = 0 \ , \label{eq:undeformed-eom-2} \\
  E^\Gamma_\alpha & = & \bar \cD \cH_{\alpha \beta} \Gamma^\beta +
  \bar \cD (\cH_{\alpha \betab} \Gammab^\betab) = 0
  \ , \label{eq:undeformed-eom-3} \\
  E^\Gammab_\alphab & = & \cD (\cH_{\beta \alphab} \Gamma^\beta) - \cD
  \cH_{\alphab \betab} \Gammab^\betab = 0 \ . \label{eq:undeformed-eom-4}
\end{eqnarray}

\section{Marginal deformations} \label{sec:marginal-deformations}

In this section we consider the marginal deformations to the lowest
order in $\alpha'$ and first order in conformal perturbation theory.

It can be seen that much like in the four-dimensional case
\cite{Green:2010da} there are no K\"ahler deformations of the form
\begin{displaymath}
  \int d^2 z \cD \bar \cD X \ .
\end{displaymath}
The only type of marginal deformations are of the form
\begin{equation}
  S_\cW = - \frac{i}{8 \pi \alpha'} \int d^2 z \, \cD \cW +
  \mathrm{h.c.} \ ,
\end{equation}
where $\cW$ must be a chiral primary of weights (1, $\frac{1}{2}$) and
$\mathrm{U}(1)_\mathrm{R}$ charge +1. (For $S_\cW$ to be truly marginal, these
conditions should hold to any order in conformal perturbation theory
and in $\alpha'$.)

Since $\cD^2 = 0$, $S_\cW$ clearly remains unmodified under $\cW \to
\cW + \cD Y$, where $Y$ has conformal weight (1, 0) and R-charge
+2. In the absence of world-sheet boundaries, it is also invariant
under $\cW \to \cW + \p \cZ$ with $\cZ$ being a superfield of weight
(0, $\frac{1}{2})$ and R-charge +1, and $\p \cZ$ is required to be
chiral.  Finally, if we deform using $\cW' = \cW + \bar \cD X$ one
obtains an equivalent deformation $S_{\cW'}$ because
\begin{equation}
  S_{\cW'} - S_\cW = \int d^2 z \, \cD \bar \cD X ,
\end{equation}
which is a trivial deformation.

A short note about the condition of chirality is in order. Working in
conformal perturbation theory, we should expand the deformed action
around the undeformed conformal theory. Therefore, we should treat the
deformation $S_\cW$ as a series of operator insertions in the
undeformed correlation function evaluated at the conformal
point. Insertions in a path integral satisfy the equations of motion
of the undeformed action (up to possible contact terms with other
insertions). Another point of view is that terms in the action that
are proportional to the equations of motion can be removed by a field
redefinition.  Henceforth, \emph{on-shell} will always mean on-shell
with respect to the undeformed equations of motion.

Since our analysis is done at the first order in conformal
perturbation theory and at tree-level in $\alpha'$, all the fields have their
classical dimensions and we can treat the deformation as a classical
object. The most general deformation with the required (1,1) conformal
weight and R-charge +1 is
\begin{eqnarray} \label{eq:marginal-deformation}
  \cW & = & (\Lambda^\alphab_{\ib \beta} (\Phi, \Phib) \Gamma_\alphab
  \Gamma^\beta + \Lambda^\alpha_{\ib \beta} \Gammab_\alpha \Gamma^\beta +
  \Lambda^\alpha_{\ib \betab}(\Phi, \Phib) \Gammab_\alpha
  \Gammab^\betab) \bar \cD \Phi^\ib + \nonumber \\
  && {} + (Y_{i \jb} (\Phi, \Phib) \p
  \Phi^i + g_{i \kb} Z_\jb{}^i(\Phi, \Phib) \p \Phib^\kb) \bar \cD
  \Phib^\jb \ ,
\end{eqnarray}
where $\Lambda^\alphab_{\ib \beta} = \cH^{\alphab \gamma} \Lambda_{\ib
  \gamma \beta}$ ($\Lambda_{\ib \alpha \beta} = -\Lambda_{\ib \beta
  \alpha}$) and $\Lambda^\alpha_{\ib \betab} = \cH^{\gammab \alpha}
\Lambda_{\ib \gammab \betab}$ ($\Lambda_{\ib \alphab \betab} = -
\Lambda_{\ib \betab \alphab}$). A term of the form
\begin{displaymath}
  \tilde \cW = (\Omega_{\alpha \beta}(\Phi, \Phib) \Gamma^\alpha +
  \Omega_{\alphab \betab}(\Phi, \Phib) \Gammab^\alphab) (\bar \cD
  \Gammab^\betab + \bar \cA^\betab_{\ib  \gammab} \bar \cD \Phib^\ib
  \Gammab^\gammab) 
\end{displaymath}
(where the derivative has been replaced by a gauge-covariant
derivative to maintain gauge-invariance) does not appear because it
can be absorbed in the deformation (\ref{eq:marginal-deformation}) by
using the undeformed equations of motion.

For the theory to be well defined on the entire compact space, the
deformation parameters must be sections of the appropriate bundles:
\begin{displaymath}
  \Lambda \in \Gamma(\Omega^{0,1} \otimes \mathrm{End}\,V) \ , \quad
  Y \in \Gamma(\Omega^{1,1}) \ , \quad
  Z \in \Gamma(\Omega^{0,1} \otimes \mathrm{T}^{1,0} M) \ .
\end{displaymath}

The deformation (\ref{eq:marginal-deformation}) is not manifestly
chiral as it depends on  anti-chiral fields as well as chiral
ones. However, as discussed above, it needs only be chiral on-shell in
order to preserve (0,2) supersymmetry in conformal perturbation
theory. On-shell
\begin{eqnarray*}
  \bar \cD \cW & = & \left( \p_\jb \Lambda^\alphab_{\ib \beta} + \bar
  \cA^\alphab_{\jb \gammab} \Lambda^\gammab_{\ib \beta} + \frac{i}{2}
  Z_\ib{}^k \tilde \cF^\alphab_{k \jb \beta} \right) \Gamma_\alphab
  \Gamma^\beta \bar \cD \Phib^\jb \bar \cD \Phib^\ib + \\
  && {} + \left( \p_\jb \Lambda^\alpha_{\jb \beta} -
  \Lambda^\alpha_{\ib \gammab} \bar \cA^\gammab_{\jb \beta} -
  \Lambda^\alpha_{\ib \gammab} \bar \cA^\gammab_{\jb \beta} - i
  Z_\ib{}^k \tilde \cF^\alpha_{\ib k \beta} \right) \Gammab_\alpha
  \Gamma^\beta \bar \cD \Phib^\jb \bar \cD \Phib^\ib + \\
  && {} + \left( \p_\jb \Lambda^\alpha_{\ib \betab} - \bar
  \cA^\gammab_{\jb \betab} \Lambda^\alpha_{\ib \gammab} + \frac{i}{2}
  Z_\jb{}^k \tilde \cF^\alpha_{\ib k \betab} \right) \Gammab_\alpha
  \Gammab^\betab \bar \cD \Phib^\jb \bar \cD \Phib^\ib + \\
  && {} + (\p_\kb Y_{i \jb} + Z_\jb{}^l H_{i \kb l}) \p \Phi^i \bar
  \cD \Phib^\jb \bar \cD \Phi^\kb + g_{i \kb} \p_\lb Z_\jb{}^i \p
  \Phib^\kb \bar \cD \Phib^\lb \bar \cD \Phib^\jb \ .
\end{eqnarray*}
Requiring that $\bar \cD \cW = 0$ yields the following constraints of
the deformation parameters
\begin{eqnarray} \label{eq:deformation-constraints}
  \cH_{\alpha \gammab} \left( \p_{[ \jb} \Lambda^\gammab_{\ib] \beta} + \bar
  \cA^\gammab_{[\jb \deltab} \Lambda^\deltab_{\ib] \beta} + \frac{i}{2}
  Z_{[\ib}{}^k \tilde \cF^\gammab_{|k| \jb] \beta} \right) - (\alpha
  \leftrightarrow \beta) & = & 0 \ , \nonumber \\ 
  \p_{[\jb} \Lambda^\alpha_{\ib] \beta} - \Lambda^\alpha_{[\ib \gammab} \bar
    \cA^\gammab_{\jb] \beta} - i Z_{[\ib}{}^k \tilde \cF^\alpha_{\jb]
    k \beta} & = & 0 \ , \nonumber \\
  \cH_{\gamma \alphab} \left( \p_{[\jb} \Lambda^\gamma_{\ib] \betab} - \bar
  \cA^\deltab_{[\jb \betab} \Lambda^\gamma_{\ib] \deltab} + \frac{i}{2}
  Z_{[\jb}{}^k \tilde \cF^\gamma_{\ib] k \betab} \right) - (\alphab
  \leftrightarrow \betab) & = & 0 \ , \nonumber \\  
  \p_{[\kb} Y_{|i|\jb]} - H_{i[\kb|l|} Z_{\jb]}{}^l & = & 0 \ ,
  \nonumber \\
  \p_{[\lb} Z_{\jb]}{}^i & = & 0 \ ,
\end{eqnarray}
where $[\dots]$ denotes anti-symmetrization with respect to space
indices only and indices between bars are excluded from the
anti-symmetrization.

As discussed earlier, deformations are subject to the equivalence relation
$\cW \simeq \cW + \bar \cD X + \p \cZ$. The most general $X$ of weight
(1,0) and R-charge 0 is (again at the classical level)
\begin{eqnarray}
  X & = & \lambda_{\alpha \beta}(\Phi, \Phib) \Gamma^\alpha \Gamma^\beta +
  \lambda^\alpha{}_\beta(\Phi, \Phib) \Gammab_\alpha \Gamma^\beta +
  \lambda_{\alphab \betab}(\Phi, \Phib) \Gammab^\alphab \Gammab^\betab
  + \nonumber \\
  && {} + \mu_i(\Phi, \Phib) \p \Phi^i + g_{i\jb} \zeta^i(\Phi, \Phib) \p
  \Phib^\jb \ ,
\end{eqnarray}
where
\begin{displaymath}
  \lambda^\alpha{}_\beta, \lambda^\alpha{}_\betab,
  \lambda^\alphab{}_\beta \in \Gamma(\mathrm{End}\,V) \ , \quad
  \mu \in \Gamma(\Omega^{1,0}(M)) \ , \quad
  \zeta \in \Gamma(\mathrm{T}^{1,0}M) \ .
\end{displaymath}
The most general $\cZ$ of weight (0,$\frac{1}{2}$) and R-charge +1 is
\begin{equation}
  \cZ = \xi_\jb(\Phi, \Phib) \bar \cD \Phib^\jb \ .
\end{equation}
$\p \cZ$ is chiral on-shell provided
\begin{eqnarray} \label{eq:Z-constriants}
  \nabla^-_k \p_{[\jb} \xi_{\ib]} = 0 \ , \quad
  \nabla^-_\kb \p_{[\jb} \xi_{\ib]} & = & 0 \ , \nonumber \\
  g^{\ib j} (\p_{[\ib} \xi_{\lb]} \tilde \cF^\alphab_{j \kb \beta} -
  \p_{[\ib} \xi_{\kb]} \tilde \cF^\alphab_{\lb j \beta}) & = & 0 \ ,
  \nonumber \\
  g^{\ib j} ( \p_{[\ib} \xi_{\lb]} \tilde \cF^\alpha_{\kb j \beta} -
  \p_{[\ib} \xi_{\kb]} \tilde \cF^\alpha_{\kb j \beta}) & = & 0 \ ,
  \nonumber \\
  g^{\ib j} (\p_{[\ib} \xi_{\lb]} \tilde \cF^\alpha_{\kb j \betab} -
  \p_{[\ib} \xi_{\kb]} \tilde \cF^\alpha_{\lb j \betab}) & = &0 \ .
\end{eqnarray}
(The relations \cite{Strominger:1986uh}
\begin{eqnarray}
  \Gamma_{k i \jb} & = & \frac{1}{2} H_{k i \jb} = -\frac{1}{2}(\p_k
  g_{i \jb} - \p_i g_{k \jb}) \ , \\
  \Gamma_{l \ib \jb} & = & \frac{1}{2} (\p_\ib g_{l \jb} + \p_\jb g_{l
  \jb}) \ , \quad
  H_{l \ib \jb} = \p_\ib g_{l \jb} - \p_\jb g_{l \ib}
\end{eqnarray}
were used to rewrite the result in terms of the $H$-twisted connection
$\nabla^-$.)

Putting all these together, the equivalence relation $\cW \simeq \cW +
\bar \cD X + \p \cZ$ in component form are
\begin{eqnarray}
  \Lambda^\alphab_{\ib \beta} & \simeq & \Lambda^\alphab_{\ib \beta} +
  \cH^{\alphab \gamma} \p_\ib \lambda_{\gamma \beta} + \bar
  \cA^\alphab_{\ib \gamma} \lambda^\gamma{}_\beta + \frac{i}{2}
  (\zeta^j + g^{\kb j} \xi_\kb) \tilde \cF^\alphab_{j \ib \beta} \ ,
  \\
  \Lambda^\alpha_{\ib \beta} & \simeq & \Lambda^\alpha_{\ib \beta} +
  \p_\ib \lambda^\alpha{}_\beta - 2 \lambda^\alpha{}_\gammab \bar
  \cA^\gammab_{\ib \beta} - i (\zeta^j + g^{\kb j} \xi_\kb) \tilde
  \cF^\alpha_{\ib j \beta} \ , \\
  \Lambda^\alpha_{\ib \betab} & \simeq & \Lambda^\alpha_{\ib \betab} +
  \p_\ib \lambda^\alpha{}_\betab - \bar \cA^\gammab_{\ib \betab}
  \lambda^\alpha{}_\gammab - \frac{i}{2} (\zeta^j + g^{\kb j} \xi_\kb)
  \tilde \cF^\alpha_{\ib j \betab} \ , \\
  Y_{i \jb} & \simeq & Y_{i \jb} + \p_\jb \mu_i + \p_i \xi_\jb + H_{i
    \jb k} (\zeta^k + g^{\lb k} \xi_\lb) \ , \\
  Z_\jb{}^i & \simeq & Z_\jb{}^i + \p_\jb (\zeta^i + g^{\kb i}
  \xi_\kb) + g^{\kb i} (\p_\kb \xi_\jb - \p_\jb \xi_\kb)
  \ . \label{eq:equivalence-relations}
\end{eqnarray}

\section{An example: $G = G' \times \mathrm{E}_8$,
  $\mathrm{U}(1)_\mathrm{L} \subset G'$} \label{sec:examples}

In this section we reconsider the case in which the bundle's surviving
structure group is $G = G' \times \mathrm{E}_8$ with $G'$ containing a
$\mathrm{U}(1)_\mathrm{L}$ factor \cite{Melnikov:2011ez}. The analysis
here differs from that in \cite{Melnikov:2011ez} by the inclusion of
bundle deformations which break the $\mathrm{U}(1)_\mathrm{L}$
symmetry.

In this case
\begin{equation}
  \cH_{\alpha \beta} = \cH_{\alphab \betab} = 0 \ ,
\end{equation}
from which it follows that
\begin{equation}
  \cA^\alpha_{i \betab} = \bar \cA^\alphab_{\ib \beta} = 0 \ , \quad
  \tilde \cF^\alphab_{j \kb \beta} = \tilde \cF^\alpha_{\kb j \betab}
  = 0 , \quad \tilde \cF^\alpha_{\kb j \beta} = \cF^\alpha_{\kb j
    \beta} \ .
\end{equation}
Thus, the constraints on the deformations
(\ref{eq:deformation-constraints}) become
\begin{eqnarray}
  \cH_{\alpha \gammab} (\p_{[\jb} \Lambda^\gammab_{\ib] \beta} + \bar
  \cA^\gammab_{[\jb \deltab} \Lambda^\deltab_{\ib]\beta}) - (\alpha
  \leftrightarrow \beta) & = & 0 \ , \nonumber \\
  \p_{[ \jb} \Lambda^\alpha_{\ib] \beta} - i Z_{[\ib}{}^k
    \cF^\alpha_{\jb] k \beta} & = & 0 \ , \nonumber \\
  \cH_{\gamma \alphab} (\p_{[\jb} \Lambda^\gamma_{\ib] \betab} - \bar
  \cA^\deltab_{[\jb \betab} \Lambda^\gamma_{\ib] \deltab}) - (\alpha
  \leftrightarrow \beta) & = & 0 \ , \nonumber \\
  \p_\kb Y_{i \jb} - H_{i \kb l} Z_\jb{}^l & = & \p_\jb Y_{i \kb} - H_{i
    \jb l} Z_\kb{}^l \ , \nonumber \\
  \p_{[\kb} Z_{\jb]}{}^i & = & 0 \ .
\end{eqnarray}
The components of $\cZ$ should satisfy
\begin{eqnarray}
  \nabla^-_k \p_{[\ib} \xi_{\jb]} & = & 0 \ , \nonumber \\
  \nabla^-_\kb \p_{[\ib} \xi_{\jb]} & = & 0 \ , \nonumber \\
  g^{\ib j} (\p_{[\ib} \xi_{\mb]} \cF^\alpha_{\kb j \beta} - \p_{[ib}
    \xi_{\kb]} \cF^\alpha_{\mb j \beta}) & = & 0 \ .
\end{eqnarray}
The equivalence relations (\ref{eq:equivalence-relations}) are then
reduced to
\begin{eqnarray}
  \Lambda^\alphab_{\ib \beta} & \simeq & \Lambda^\alphab_{\ib \beta} +
  \cH^{\alphab \gamma} \p_\ib \lambda_{\gamma \beta}
  \ , \label{eq:extra-deformation-equiv-1} \\ 
  \Lambda^\alpha_{\ib \beta} & \simeq & \Lambda^\alpha_{\ib \beta} +
  \p_\ib \lambda^\alpha{}_\beta - i (\zeta^j +g^{\kb j} \xi_\kb)
  \cF^\alpha_{\ib j \beta} \ , \label{eq:EndV-equivalence} \\
  \Lambda^\alpha_{\ib \betab} & \simeq & \Lambda^\alpha_{\ib \betab}
  + \p_\ib \lambda^\alpha{}_\betab - \bar \cA^\gammab_{\ib \betab}
  \lambda^\alpha{}_\gammab \ , \label{eq:extra-deformation-equiv-2} \\
  Y_{i \jb} & \simeq & Y_{i \jb} + \p_\jb \mu_i + \p_i \xi_\jb + H_{i
    \jb k} (\zeta^k + g^{\lb k} \xi_\lb) \ , \label{eq:Kahler-deformation-equiv} \\
  Z_\jb{}^i & \simeq & Z_\jb{}^i + \p_\jb (\zeta^i + g^{\kb i}
  \xi_\kb) + g^{\kb i} (\p_\kb \xi_\jb - \p_\jb \xi_\kb) \ . \label{eq:cx-equiv}
\end{eqnarray}
These are the same as the results obtained in \cite{Melnikov:2011ez}
with the addition of deformations which break the $\mathrm{U}(1)_\mathrm{L}$
symmetry.

We can bring the extra deformation parameterized by
$\Lambda^\alphab_{\ib \beta}$ and $\Lambda^\alpha_{\ib \betab}$ to a
nicer form. Doing a little algebra yields
\begin{eqnarray}
  \cH_{\alpha \gammab} \p_\jb \Lambda^\gammab_{\ib \beta} & = &
  \p_\jb \Lambda_{\ib \alpha \beta} - \p_\jb \cH_{\alpha \gammab}
  \Lambda^\gammab_{\ib \beta} \ , \nonumber \\
  \cH_{\alpha \gammab} \bar \cA^\gammab_{\jb \deltab}
  \Gamma^\deltab_{\ib \beta} & = & \p_\jb \cH_{\alpha \gammab}
  \Lambda^\gammab_{\ib \beta} \ ,
\end{eqnarray}
so we can rewrite (\ref{eq:extra-deformation-equiv-1}) and its associated
equivalence relation as
\begin{equation}
  \p_{[\jb} \Lambda_{\ib] \alpha \beta} = 0 \ , \quad
  \Lambda_{\ib \alpha \beta} \simeq \Lambda_{\ib \alpha \beta} +
  \p_\ib \lambda_{\alpha \beta} \ .
\end{equation}
Hence, these extra deformations are elements of $H^1 (M, V^* \wedge
V^*)$. A similar manipulation of (\ref{eq:extra-deformation-equiv-2}) gives
\begin{equation}
  \p_{[\ib} \Lambda^{\alpha \beta}_{\jb]} = 0 \ , \quad
  \Lambda^{\alpha \beta}_\ib \simeq \Lambda^{\alpha \beta} + \p_\ib
  \lambda^{\alpha \beta},
\end{equation}
which are elements of $H^1 (M, V \wedge V)$. 

In particular, for the heterotic string compactified on a Calabi-Yau
three-fold and the standard embedding of the tangent bundle in the
vector bundle, which breaks the first $\mathrm{E}_8$ into
$\mathrm{E}_6$, these new deformations will be in $H^1(M,\mathrm{T}^*M
\wedge \mathrm{T}^*M) \simeq H^{2,1}(M)$ and $H^1(M, \mathrm{T}M
\wedge \mathrm{T}M) \simeq H^{1,1}(M)$.

These are the same as the results obtained in \cite{Melnikov:2011ez}
with the addition of deformations that break the
$\mathrm{U}(1)_\mathrm{L}$ symmetry. To our knowledge, these
deformations are new.\footnote{The author would like to thank
  I.~V.~Melnikov for discussions on this point as well as for sharing
  some results obtained by him and by J.~McOrist.} A possible
application of these new deformations is breaking the $\mathrm{E}_6$
gauge group of the standard embedding to $\mathrm{SO}(10)$. In this
case the $\mathbf{78}$ adjoint representation of $\mathrm{E}_6$ is
decomposed under its $\mathrm{SO}(10) \times \mathrm{U}(1)_\mathrm{L}$
into $\mathbf{45}_0 \oplus \mathbf{16}_{-3} \oplus
\overline{\mathbf{16}}_3 \oplus \mathbf{1}_0$ \cite{Slansky:1981yr}. A
deformation breaking the $\mathrm{U}(1)_\mathrm{L}$ should Higgs all
but the $\mathbf{45}$ of $\mathrm{SO}(10)$. The $\mathbf{1}$ is
clearly lifted and the two spinor representations must become massive
as well for the consistency of the low-energy effective theory.

\textbf{Acknowledgments} The author would like to thank Ilarion
Melnikov and Ronen Plesser for collaboration on a related but yet
unpublished project on marginal deformations of (0,2) SCFTs, whose
intermediate results proved to be useful for this paper. The author
would also like to thank Ilarion Melnikov for many useful discussions,
sharing some of his notes and for commenting on an initial draft.
This research was supported by FAPESP fellowship 2010/07439-0.

\bibliography{ido-references.bib}
\bibliographystyle{amsplain}
\end{document}